\documentclass[aps,prl,twocolumn,showpacs,floatfix]{revtex4-1}

\usepackage{amsmath}
\usepackage{amssymb}
\usepackage{graphicx}

\usepackage{color}

\begin{document}

\title{Accounting for the kinetics in order parameter analysis: lessons
from theoretical models and a disordered peptide}

\date{\today}

\author{Ganna Berezovska}
\author{Diego Prada-Gracia}
\author{Stefano Mostarda}
\author{Francesco Rao}
\email{francesco.rao@frias.uni-freiburg.de}

\affiliation{Freiburg Institute for Advanced Studies, School of Soft Matter
Research, Freiburg im Breisgau, Germany.}

\begin{abstract}

  Molecular simulations as well as single molecule experiments have been widely
  analyzed in terms of order parameters, the latter representing candidate
  probes for the relevant degrees of freedom. Notwithstanding this approach is
  very intuitive, mounting evidence showed that such description is not
  accurate, leading to ambiguous definitions of states and wrong kinetics. To
  overcome these limitations a framework making use of order parameter
  fluctuations in conjunction with complex network analysis is investigated.
  Derived from recent advances in the analysis of single molecule time traces,
  this approach takes into account of the fluctuations around each time point
  to distinguish between states that have similar values of the order parameter
  but different dynamics.  Snapshots with similar fluctuations are used as
  nodes of a transition network, the clusterization of which into states
  provides accurate Markov-State-Models of the system under study.  Application
  of the methodology to theoretical models with a noisy order parameter as well
  as the dynamics of a disordered peptide illustrates the possibility to build
  accurate descriptions of molecular processes on the sole basis of order
  parameter time series without using any supplementary information.

\end{abstract}

\maketitle

\section{Introduction}

Order parameters are conventionally used for the characterization of complex
molecular processes \cite{du1998, benkovic2008}.  Inter-atomic distances or a
combination of them are common choices, providing an intuitive description in
terms of free-energy projections \cite{lazaridis1997, dinner2000, zhou2001,
rao2003, cecchini2004}.  Unfortunately, it has been repeatedly found that
reduced descriptions based on order parameters are often inaccurate
\cite{rao2004, krivov2004, rao2005, hegger2007, muff2008, maisuradze2009}. The
origin of the failure is due to overlaps in the order parameter distribution,
i.e., configurations with different properties corresponding to the same value
of the coordinate, making the discrimination between states ambiguous
\cite{rao2005,muff2008}. From the point of view of the dynamics, spurious
recrossings at the borders result in lower free-energy barriers and
artificially faster kinetics \cite{krivov2011}.

To improve on this situation, a new arsenal of tools emerged making use of
complex networks and the theory of stochastic processes.
Configuration-space-networks, referred as Markov-State-Models when the Markov
property is satisfied, provide high resolution free-energy landscapes of
complex molecular processes \cite{rao2004, krivov2004, gfeller2007, noe2007,
chodera2007, prada2009, rao2010}.  The main idea behind this approach is to map
the molecular dynamics onto a transition network. Nodes and links represent
sampled system configurations (\emph{microstates}) and the transitions between
them as observed in the molecular dynamics, respectively.  The resulting
transition network stores the entire kinetical information in the form of link
weights and node connectivity, providing a compact representation of the
molecular trajectory.  Within this approach, free-energy representations are
obtained in a more universal way without using arbitrarily projections on order
parameters.  Both thermodynamics and kinetics come from the analysis of the
transition network with methods like network clusterization algorithms
\cite{gfeller2007, prada2009, rao2010jpcl}, network flow analysis
\cite{krivov2004, krivov2006, rao2010} and spectral methods \cite{noe2007,
chodera2007}.

\begin{figure}
\includegraphics[width=80mm] {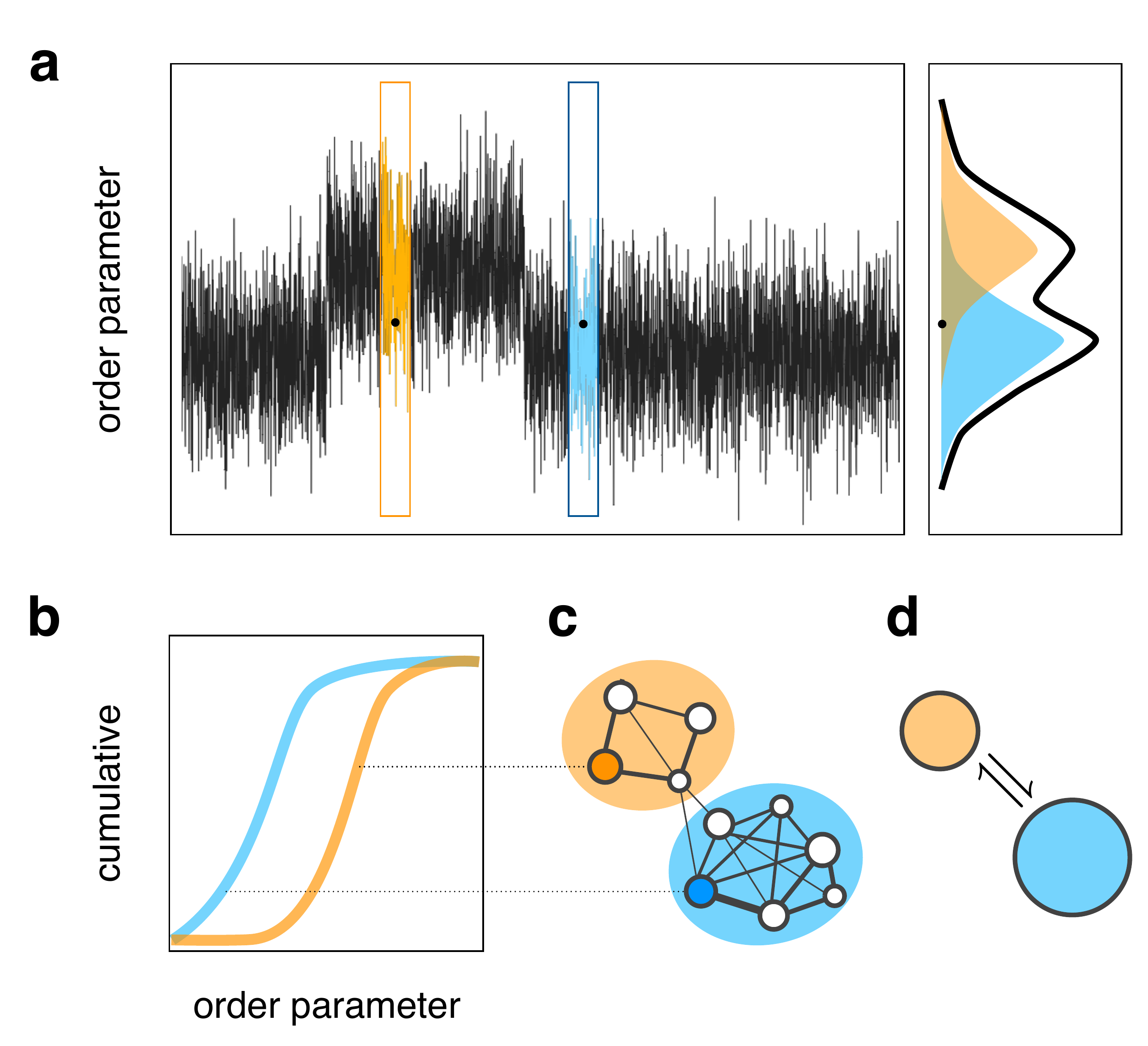}
  \caption{Local-fluctuations order parameter analysis. (a) The time series of
    an order parameter and its distribution (black lines).  Two snapshots with
    the same value of the order parameter but belonging to different states are
    characterized by distinct local distributions (orange and light blue,
    respectively). (b)  A Kolmogorov-Smirnov test evaluates the similarity of
    the cumulative of the two distributions.  Snapshots with similar
    distributions belong to the same \emph{microstate}.  (c) Microstates and
    the transitions between them  represent  nodes and links of a
    configuration-space-network, respectively.  Network clusterization
    techniques allow the lumping of kinetically homogeneous regions of the
  network into states (orange and light blue areas) (d) States are used to
build a reduced Markov-State-Model of the original molecular process (see the Theory
section for further details).}
  \label{fig:strategy}
\end{figure}

Besides the advantages, a general strategy towards microstates building is
still missing, making the initial mapping of the molecular trajectory onto a
network non-trivial.  Even for the well-studied case of structured peptides and
the folding of small proteins, there is no consensus on the best practice
\cite{rao2010, hua2012}. Moreover, a broad set of problems including molecular
association \cite{buch2011, huang2011}, the analysis of intrinsically
disordered proteins \cite{uversky2008, knott2012} and liquids
\cite{errington2001, yan2004} are very hard to tackle with the current
methodology. As shown for the case of water, \textit{ad-hoc} strategies are
needed \cite{rao2010jpcb, garrett2011, prada2012}.  Ironically, many of these
processes can be qualitatively described  by the analysis of conventional order
parameters. 

In this work, an effort is made to reconcile the intuitive aspect of order
parameters with the predictive power of transition networks, overcoming some of
the limitations of both methodologies.  The strategy couples a recently
developed framework for the analysis of single molecule traces \cite{baba2007,
schuetz2010, baba2011} using network clusterization techniques in order to
obtain accurate kinetic models from conventional order parameter time series.
Applications to theoretical models and molecular dynamics simulations of a
disordered peptide are presented. Our results suggest a general approach to
analyze molecular processes with high accuracy on the sole basis of
conventional order parameter time series.

\section{Theory}

\subsection{Configuration-space networks from conventional order parameters}

{\it General motivation.} Order parameters allow for intuitive descriptions of
molecular processes. Unfortunately, such descriptions can be highly inaccurate
due to the presence of \emph{overlaps}, i.e., configurations with different
properties corresponding to the same value of the order parameter
\cite{rao2005,muff2008}. An important improvement in this respect was the
introduction of configuration-space-networks, providing accurate and concise
descriptions of the system kinetics and thermodynamics \cite{rao2004,
krivov2004, gfeller2007, noe2007, chodera2007, prada2009, rao2010}. Their
application however is still limited, lacking a general way to build the
transition network.

To overcome this impasse, a potential strategy makes use of a recently
introduced framework for the analysis of single molecule experiments
\cite{baba2007, schuetz2010, baba2011}.  Local fluctuations of a given
coordinate bring relevant information on the underlying free-energy surface.
That is, two points with similar values of the coordinate but belonging to
different states are characterized by distinct local distributions (see orange
and blue areas in Fig.~\ref{fig:strategy}a).  In order to characterize the
molecular process, this information can be used in different ways, going from
the concept of state ``candidate" based on escape times
\cite{baba2007,baba2011} to cut-based free-energy profiles \cite{schuetz2010}.
The latter approach proposed a way to build the system microstates based on the
local fluctuations of an arbitrary inter-atomic distance, showing that the
folding barrier and native state population of a small protein are correctly
recovered.  On the other hand, the limited information contained in a single
distance made the full reconstruction of the unfolded state hard.

Here, local fluctuations are exploited towards a better characterization of
order parameter time series, overcoming the problems raised by the presence of
overlaps.  The proposed protocol works as follows: (i) based on the time series
of the order parameter a set of microstates is constructed; (ii) the resulting
configuration-space-network is built; (iii) the presence of states is found by
performing a network clusterization algorithm; (iv) a reduced kinetic model is
built based on the found states.  These four steps are described in detail
below.

{\it Microstate building.} The microstates accounting for the local
fluctuations of the order parameter were built as suggested in
Ref.~\cite{schuetz2010}.  As such, each time point of the trajectory $t_{i}$
was associated with a corresponding time window $[t_{i}-\tau/2,t_{i}+\tau/2]$.
Two time points were considered to be \emph{similar} if they have comparable
distributions of the order parameter.  Snapshot similarity, $D$, was estimated
by comparing the cumulative of the two distributions via a Kolmogorov-Smirnov
test \cite{KST} which checks whether two samples belong to the same
distribution or not (Fig.~\ref{fig:strategy}b). $D$ was defined as the maximum
difference of the two cumulative distributions. Two samples belong to the same
distribution, and thus to the same microstate,  if the condition $ D\leq \zeta
\sqrt{2/\tau} $ was fulfilled.  The acceptance cutoff $\zeta$ corresponds to a
certain confidence level. Being $\tau$ and $\zeta$ related, we fixed the latter
value to 0.5 and let $\tau$ vary. Comparisons were made along the trajectory
using the leader algorithm in a way that each time point was associated to
a microstate at the end of the procedure \cite{seeber2007, schuetz2010}.

{\it The configuration-space-network. } The resulting time series of
microstates was mapped onto a configuration-space-network
(Fig.~\ref{fig:strategy}c).  Microstates represent network nodes and a link
between them exists if they were successively visited along the molecular
trajectory. For each link detailed balance was imposed by making an average of
the number of transitions in both directions.

{\it Network clusterization.} A clusterization algorithm was applied for the
analysis of the configuration-space-network in order to detect the presence of
states. In fact, free-energy basins are represented as densely connected
regions of the configuration-space-network \cite{gfeller2007}. It was shown
\cite{gfeller2007, prada2009} that those regions can be automatically detected
by using the Markov-Clustering-Algorithm (MCL) for network clusterization
\cite{enright2002}.  This approach is based on the evolution of random walkers
on the network, resulting in a kinetically accurate network splitting. Hence,
dynamical interconversions within a cluster are faster than transitions to
other regions of the network. Being separated by barriers, network clusters
represent free-energy basins (i.e., the \emph{states}) of the system (orange
and blue areas in Fig.~\ref{fig:strategy}c). A parameter $p>1$ tunes the
granularity of the clusterization. Larger values of the parameter (e.g.
$p>1.5$) result in an increased number of clusters while the most relevant
states (i.e. separated by the highest barriers) are already detected with $p$
between 1.2 and 1.4 \cite{gfeller2007, prada2009}.

{\it Reduced kinetic model.} The most populated clusters were used as states of
a reduced Markov-State-Model. Transition probabilities were estimated from the
original transition network by summing up all the links connecting any two
states (orange and blue areas in Fig.~\ref{fig:strategy}c-d). It is worth
noting that while this reduced kinetic model satisfies the Markov property
\cite{rao2010}, this is not generally the case for the starting transition
network. (This property is in any case not needed when it comes to network
clusterization \cite{gfeller2007, prada2009}.) The accuracy of the kinetic
model was estimated with the use of first-passage-time distributions.

\section{Methods}

\subsection{Two-state model} 

A stochastic two-state model with transition probability $p_{ij}=0.01$ was
built.  The latter probability completely defines the kinetics of the model.
The time evolution was monitored by an artificially defined order parameter $Q$
in a way that $Q$ cannot distinguish between the two states unambiguously (see
dark blue line in Fig.~\ref{fig:model}a for a sample time series).  $Q$ was
associated to an energy function $U_1=\alpha Q^2$ and $U_2=\alpha (Q-1)^2$ for
the two states, respectively. As such, the first and second states
preferentially visit different values of the order parameter.  Within each
state, the time evolution of $Q$ was obtained by a conventional Metropolis
criterion $\min\left[1,\exp\left(-\beta\Delta U\right)\right]$ with $\beta$
regulating the amount of fluctuations. Choosing $\alpha=16$, values of $\beta$
close to 1 suppress fluctuations while smaller values enhance them (most of the
treatment below was done for the case of large fluctuations with $\beta=0.3$).
This model was used to generate order parameter time series of $10^5$ steps.

\subsection{Molecular dynamics simulations} 

Simulations of the (Gly-Ser)$_2$ flexible linker peptide were performed using
the all-atom CHARMM force-field version 27 \cite{brooks1983, brooks2009} as
implemented in the program ACEMD \cite{harvey2009acemd}. All calculations were
done on NVIDIA GTX680 graphics cards. The system was solvated into a box
containing 1560 TIP3P waters. After equilibration, the molecular dynamics was
run for 1$\mu$s in the NVT ensemble at 300K, using the Langevin algorithm.  An
integration time step of 4 fs was used by rescaling the hydrogen mass to 4 amu
together with mass repartitioning \cite{Feenstra1999}. The radius of gyration
was calculated with the program WORDOM \cite{seeber2007, seeber2011},
neglecting all peptide hydrogens.

\section{Results}

\begin{figure}
  \includegraphics[width=80mm]{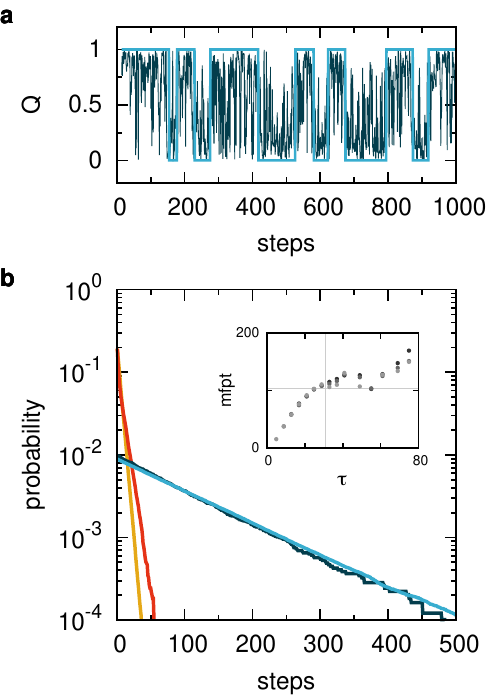} 
  \caption{Theoretical two-state model. (a) The time series of the order
    parameter $Q$ and the corresponding states obtained with the
    local-fluctuations analysis are shown in dark and light blue, respectively;
    (b) first-passage time distributions obtained by different analysis
    techniques. The distribution corresponding to the original two-state model
    is shown in dark blue. Distributions obtained from the local-fluctuations
    reduced kinetic model, a \emph{naive} two-state model and along the
    original time series using $Q=0.5$ as state separator are shown in light
    blue, yellow and red lines, respectively. The dependence of the
    mean-first-passage-time (mfpt) with the time window size $\tau$  is shown
    in the inset (MCL $p$ parameter of 1.3 and 1.4 for gray and dark gray
  points, respectively). The correct value of the mfpt and the time window value chosen for
the analysis are shown as horizontal and vertical lines, respectively.} 
  \label{fig:model}
\end{figure}

\subsection{A two-state process with large fluctuations}

The protocol described in the Theory section was applied to the time series of
a generic order parameter $Q$ with an underlying two-state dynamics. This model
served as a benchmark for the proposed network approach since the underlying
kinetics is known by construction.  The amount of fluctuations of the order
parameter $Q$  is controlled by a parameter $\beta$ (see Methods).  To
characterize the two-state behavior directly from the time series a
\emph{naive} strategy would take the distribution of $Q$, looking for the
minimum separating the two states. The value of $Q$ at the minimum, 0.5 in this
case, defines the separator between the states.  If the fluctuations around
the separator are small, $Q$ is a good order parameter in the sense that the
number of crossings of the separator represents a good estimate of the barrier
between the two states. This approach is valid for the case of small
fluctuations but breaks down as soon as the overlap between the states
increases (i.e. large fluctuations). In the latter case, most of the
contributions to the separator are coming from re-crossings: transitions
passing the separator coming back very quickly to the initial state
without reaching the end state.  The origin of recrossings is mostly due to a
sub-optimal choice of the order parameter rather than to a real physical
property.

The time series of $Q$ for the case of large fluctuations ($\beta=0.3$) was
characterized with different approaches. A particular stringent test consists
in the calculation of the first-passage-time (fpt) distribution to one of the
two states. The correct distribution is shown as a dark blue line in
Fig.~\ref{fig:model}b (mean-fpt of 103.85 steps). This distribution is greatly
influenced by the definition of the target state. If the latter is correctly
defined, the resulting distribution overlap with the one calculated from the
original two-state model. When the target state was chosen as $Q<0.5$, a
fpt distribution calculated along the trajectory resulted in a much faster
kinetics (red line, Fig.~\ref{fig:model}b). With a mfpt of 7.56 steps, the
kinetics obtained by this analysis was one order of magnitude faster with
respect to the input model.

A better description was obtained by analyzing the time series in terms of
local-fluctuations and Markov-State-Models. The microstates were obtained by
the protocol described in the Theory section (Fig.~\ref{fig:strategy}).
Clusterization of the resulting transition network resulted in the detection of
two states.  Being based on the order parameter fluctuations, these states are
not strictly separated in terms of $Q$ as shown by their time series (light
blue, Fig.~\ref{fig:model}a). These two states were used to build a new
Markov-State-Model with transition probabilities estimated from the original
transition network (see Methods).  In this case, the fpt distribution was
estimated by generating a new time series of $10^6$ steps from this model.
Strikingly, the fpt distribution calculated on this new trajectory perfectly
overlapped to the one generated by the original two-state model as shown by the
light blue line of Fig.~\ref{fig:model}b. Using a time window $\tau=30$, the
resulting mfpt is of about 114.58 steps, very close to the correct value of
103.85.  This is not the case when a two-state model was built by estimating
the transition probability from the number of times the separator $Q=0.5$ was
crossed. As expected, the fpt distribution calculated on the time series
generated by this model leads to very poor results because the inter-state
separator is dominated by recrossings (mfpt=5.2, yellow line,
Fig.~\ref{fig:model}b).  

It is important to note that the two kinetic models presented here
(corresponding to the light blue and yellow lines) were built using the same
starting information, i.e. the time series of the order parameter $Q$.
Consequently, the improvement obtained by the local-fluctuations analysis is
purely due to the different strategy applied rather than the use of
supplementary information.  

It is important to note that the predictions may depend on the choice of the
time window $\tau$ (see Theory). Caflisch and coworkers found a large range of
validity for this parameter \cite{schuetz2010}. For the present two-state
model, the time window range was evaluated against the mfpt to reach the target
state (inset of Fig.~\ref{fig:model}b).  For small time windows the value of
the mfpt was smaller with respect to the correct one (horizontal line). This is
due to the incorrect detection of states where large fluctuations were detected
as real transitions. As the value of $\tau$ was increased, the mfpt first
converged to values close to the theoretical one ($30<\tau<60$) and then
increased again. It was found that when the time window was too large, some
transitions were missed, resulting in a overall slower kinetics. Consequently,
the behavior of the mfpt as a function of $\tau$ suggests a reasonable way to
choose the time window  as the location of the first slope change ($\tau=30$,
vertical line), just before the convergence region.  Essentially identical
results were found for the MCL parameter $p=1.3-1.4$ (inset of
Fig.~\ref{fig:model}b).

\subsection{Multi-state dynamics of a disordered peptide}

\begin{figure*}[t]
  \includegraphics[width=120mm] {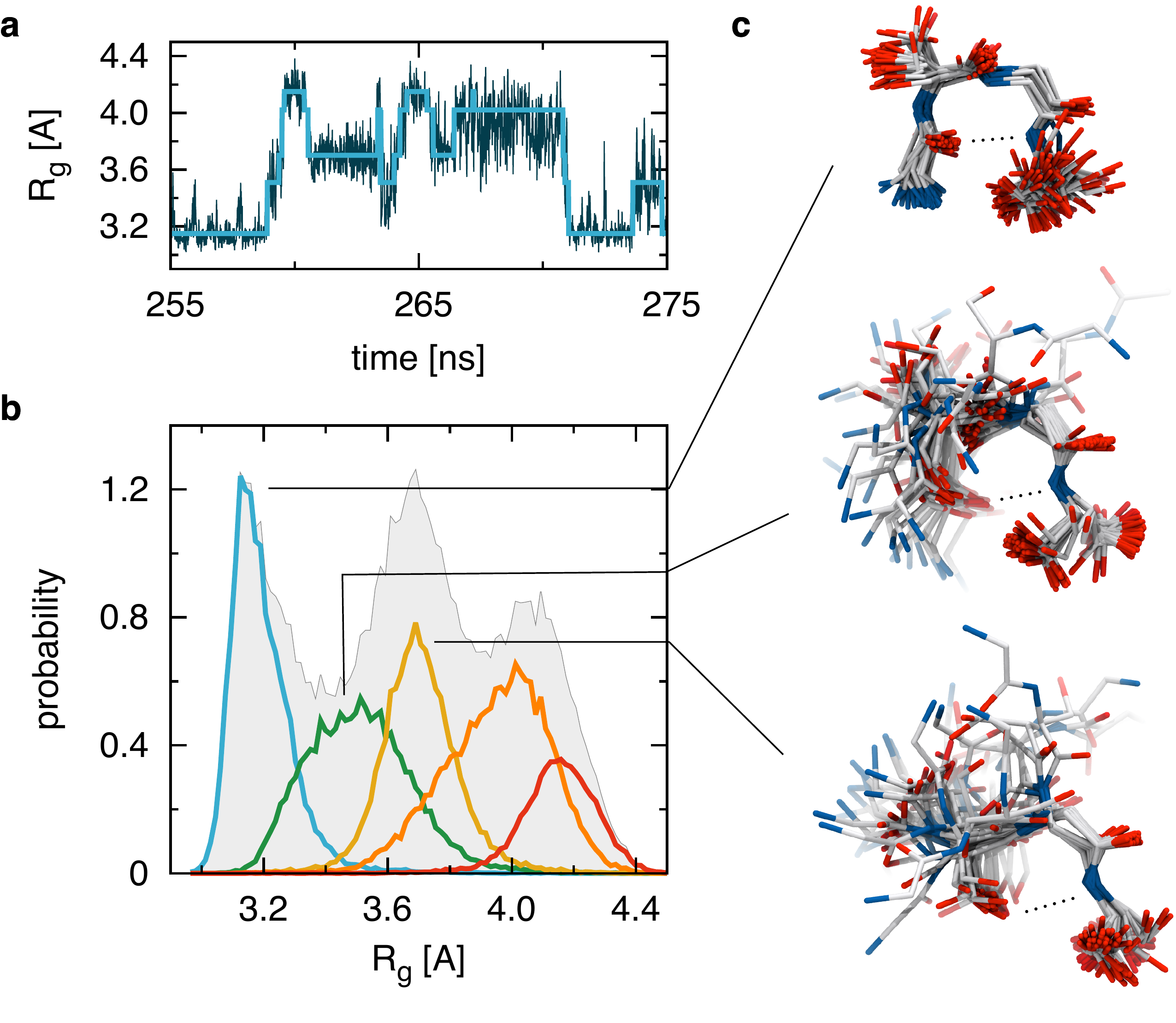}
  \caption{The (Gly-Ser)$_2$ peptide. (a) A time series stretch of the radius
    of gyration ($R_g$, dark blue) and of the detected states after the
    local-fluctuations analysis (light blue); (b) Distribution of the radius of
    gyration. The distributions from the entire trajectory and for the five
    most populated states are shown as a gray area and colored lines,
  respectively. (c) Structural characterization of the three most compact
states. For each of them, 50 random snapshots were overimposed.}
  \label{fig:gsgs}
\end{figure*}

\begin{figure}[t]
  \includegraphics[width=80mm] {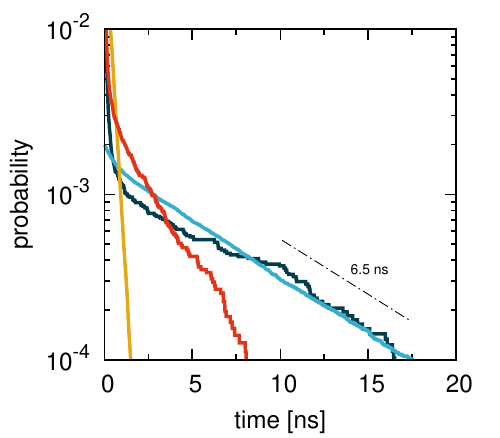}
  \caption{The (Gly-Ser)$_2$ first-passage-time distribution.  Distributions
    obtained from the local-fluctuations analysis, a \emph{naive} two-state
    model and along the original time series are shown in light blue, yellow
    and red, respectively. For the latter two cases the target of the
    relaxation was $R_g<3.4$. The relaxation kinetics to the compact state
    defined as the conformations belonging to the first peak of the $R_g$
    (\mbox{$R_g=3.16 \pm 0.01$}, see Fig.~\ref{fig:gsgs}b) is shown in dark
  blue.}
  \label{fig:gsgs-fpt}
\end{figure}

GlySer peptides have been used in experiments for quite some time as flexible
linkers \cite{bieri1999,moglich2006}.  Short stretches of this peptide are
interesting from a theoretical point of view because they are computationally
tractable, presenting non-trivial conformational disorder \cite{rao2010jpcl,
rao2011, rao2012}. In this section the local-fluctuations analysis is applied
to the dynamics of a (Gly-Ser)$_2$ peptide. To this aim, a long molecular
dynamics simulation of 1 $\mu$s was performed. It has been shown \cite{rao2012}
that the radius of gyration $R_g$ qualitatively describes the conformational
disorder of this peptide, suggesting the presence of multiple states. A time
series stretch and the distribution of the $R_g$ are shown as a dark blue line
and a gray area in panel a and b of Fig.~\ref{fig:gsgs}, respectively.

Application of the local-fluctuations analysis on the $R_g$ time series
revealed the presence of five major states ($\tau=300$ ps and $p=1.3$; these
two values were chosen following the mfpt based strategy of the previous
section).  Fig.~\ref{fig:gsgs}b shows the $R_g$ distribution of the five states
(colored lines). Interestingly, the five distributions are largely overlapping
making their detection impossible by simply looking at the total distribution.
In fact, the total distribution suggested no more than three states as
indicated by the number of peaks (gray are in Fig.~\ref{fig:gsgs}b).
Interestingly, the presence of the five states was already apparent in the raw
$R_g$ time series (dark blue line, Fig.~\ref{fig:gsgs}a). Those states became
hidden in the total distribution due to the large fluctuations, reiterating the
idea that using free-energy projections for the characterization of molecular
processes can be ambiguous.

From a structural point of view, all five states are well characterized. A
molecular representation of the three most compact states is shown in
Fig.~\ref{fig:gsgs}c. The most compact one, coded in light blue in
Fig.~\ref{fig:gsgs}b, corresponds to a loop-like structure typically found in
$\beta$-strands turns. This structure is stabilized by a hydrogen bond between
the first backbone oxygen O$_1$ and nitrogen N$_4$.  This conformation has a
population of around 23\% and $R_g \approx 3.1$ \AA.  The second state has a
population of 22\% and $R_g\approx 3.5$ \AA\ (green curve in
Fig.~\ref{fig:gsgs}b).  Its topology is very similar to the turn-like structure
but it is more disordered due to the formation of a non optimal backbone
hydrogen bond.  The third state instead is stabilized by the interaction of the
side chain oxygen of SER$_2$ with the backbone nitrogen N$_4$ (population of
20\% and $R_g\approx 3.7$ \AA, yellow curve in Fig.~\ref{fig:gsgs}b). In this
structure, the side chain substitutes the backbone oxygen as a partner in the
hydrogen bond, acting as a trap towards further compaction.  Finally, the last
two states (orange and red  curves in Fig.~\ref{fig:gsgs}b) are rather
unstructured, providing similar realizations of almost completely extended
conformations.

To check whether the kinetics of the Markov-State-Model built on top of the
five detected states reflected the same dynamics of the original trajectory, a
fpt analysis was performed (Fig.~\ref{fig:gsgs-fpt}). In light blue, the fpt
distribution to the compact state was calculated on a trajectory originated
from the Markov-State-Model.  To compare it with the molecular dynamics
simulation, the fpt to conformations with $R_g=3.16 \pm 0.01$ (the top of the
first peak in the $R_g$ distribution) was calculated along the original
molecular dynamics trajectory (dark blue line in Fig.~\ref{fig:gsgs-fpt}). This
represents a good estimate of the fpt to the compact state at long times.
Strikingly, the two curves nicely overlapped. An exponential fit of the data,
i.e. $\sim \exp(-t/t_r)$, showed a relaxation time of 6.5 ns in both cases. 

The fpt distribution calculated along the original trajectory using as target
state all conformations with $R_g<3.4$ (this is the value of the minimum of the
$R_g$ distribution) provided a much faster kinetics (red line in
Fig.~\ref{fig:gsgs-fpt}).  However, processes involving barrier crossings are
expected to have the same relaxation kinetics when either the whole state is
selected as a target or just the most probable conformation of the state (i.e.
conformations at the peak of the order parameter distribution) \cite{rao2010}.
Consequently, the  disagreement between the fpt distribution to $R_g<3.4$ and
to a more stringent target (i.e. $R_g=3.16 \pm 0.01$), provides strong evidence
that the correct estimation of the kinetics cannot be obtained directly from
the $R_g$ distribution due to the presence of recrossings.  As expected, this
becomes even worse when the Markov-State-Model is built directly from the $R_g$
distribution, i.e. by choosing as states the portions of the distribution
separated by minima and as transition probabilities the number of times the
minima were crossed (yellow line in Fig.~\ref{fig:gsgs-fpt}).

\section{Discussion}

Nowadays, molecular dynamics simulations easily produce tera scale of
data. As such, an important bottleneck in the understanding of molecular
processes is in the analysis rather then the data generation per
se. 

In this work, a strategy for the analysis of conventional order parameters time
series that is kinetics compliant was investigated.  Our results provided
strong evidence that the coupling of order parameter fluctuations with complex
network analysis represents a powerful approach to deconvolute crowded order
parameter distributions of molecular systems.  This procedure allows the
construction of kinetically accurate Markov-State-Models in a natural and
intuitive way, largely overcoming the problems raised by conventional order
parameter analysis.  Moreover, a wide range of experimentally generated time
traces coming from FRET or optical tweezers, can be readily tackled by this
methodology.

Taking into account the fluctuations within a time-window $\tau$, the approach
is able to distinguish between snapshots belonging to different states but
having the same value of the order parameter.  Towards an accurate
characterization of the kinetics the value of $\tau$ needs to be 
chosen appropriately. We proposed to estimate it on the basis of a mean-first-passage-times
analysis. Very similar in spirit to what Caflisch and collaborators proposed in
their work \cite{schuetz2010}, this procedure is more suitable to our network
clusterization approach. 

Finally, it is worth mentioning that instead of looking at better ways to
analyze conventional order parameters time series, some groups focused their
attention on the development of optimal reaction coordinates \cite{best2005,
ma2005, krivov2011numerical}. These abstract coordinates aim to correctly
characterize the molecular kinetics.  Among them, one method based on cut-based
free-energy profiles seems very promising \cite{krivov2011numerical}.  In this
approach, the coefficients of a linear combination of physical distances are
optimized against the cut-based free-energy profile. At the end of the process,
the combination which maximizes the barrier height with respect to a target
state provides the optimal reaction coordinate.  A fundamental difference
between this method and the local-fluctuations analysis is that the latter
requires the time evolution of a single (non-optimal) coordinate while the
optimization procedure makes use of a very large number of physical distances
(usually around thousands \cite{krivov2011, krivov2011numerical,steiner2012})
to perform properly.


%

\end{document}